\documentclass[prl,twocolumn,aps,floats,superscriptaddress]{revtex4}
\usepackage{graphicx,epsfig}

\def\eV{\mbox{eV}}
\def\a{\alpha}
\def\b{\beta}
\def\d{\delta}

\def\iomn{i\omega_n}
\def\inun{i\nu_n}
\def\vq{{\bf q}}
\def\vk{{\bf k}}
\def\vr{{\bf r}}
\def\vR{{\bf R}}
\def\T{\mbox{Tr}}

\def\hro{\hat{\rho}}
\def\bra{\langle}
\def\ket{\rangle}
\def\cG0{{\cal G}_0}
\def\GRRp{G^{\vR\vR'}}
\def\WRRp{W^{\vR\vR'}}
\def\GRR{G^{\vR\vR}}
\def\WRR{W^{\vR\vR}}
\def\imp{\text{imp}}

\def\cU{{\cal U}}
\def\cG{{\cal G}}
\def\GH{{G_H}}

\begin{document}

\title{First principles approach to the electronic
structure of strongly correlated systems: combining GW and DMFT}
\author{S. Biermann}
\affiliation{Laboratoire de Physique
des Solides, CNRS-UMR 8502, UPS B\^at. 510, 91405 Orsay France}
\affiliation{LPT-ENS CNRS-UMR 8549, 24 Rue
Lhomond, 75231 Paris Cedex 05, France}
\author{F. Aryasetiawan}
\affiliation{Research Institute for Computational Sciences, AIST,
1-1-1 Umezono, Tsukuba Central 2, Ibaraki 305-8568, Japan}
\author{A. Georges}
\affiliation{LPT-ENS CNRS-UMR 8549, 24 Rue
Lhomond, 75231 Paris Cedex 05, France}
\affiliation{Laboratoire de Physique
des Solides, CNRS-UMR 8502, UPS B\^at. 510, 91405 Orsay France}

\date{July,17, 2002; Revised: October, 31, 2002}
\begin{abstract}
We propose a dynamical mean field approach for calculating the
electronic structure of strongly correlated materials
from first principles.
The scheme combines the GW method with dynamical
mean field theory,
which enables one to treat strong interaction effects.
It avoids the conceptual problems inherent
to conventional ``LDA+DMFT'', such as Hubbard
interaction {\it parameters} and double counting terms.
We apply a simplified version of the approach
to the electronic structure of nickel and find encouraging
results.
\end{abstract}

\maketitle

For systems with moderate Coulomb correlations the GW method
(and its refinements)
\cite{hedin_gw,ferdi_gw_review,reining_gw_rmp}
is the tool of choice for the 
determination of excited states properties from first principles.
It is a Green's function-based method,
in which the effective screened interaction is treated at the RPA level,
and used to construct an approximation to the electronic self-energy.
This approach cures many of the artifacts encountered when the Kohn-Sham
orbitals are interpreted as physical excited states, while
they are actually auxiliary quantities within Density
Functional Theory (DFT).

Although the GW approximation (GWA)
has provided successful treatments of weakly to moderately
correlated systems such as $sp$ metals and semiconductors, applications to
more strongly correlated systems with localized orbitals
indicate a need to go beyond the GWA. For example, in ferromagnetic nickel,
it was found \cite{ferdi_gw_nickel} that the GWA is successful at predicting
the quasiparticle band-narrowing, but does not improve the (too large) exchange
splitting found in DFT calculations in the local density approximation (LDA).
The GWA also fails to reproduce the $6\eV$ satellite
observed in photoemission \cite{maartenson}.

Recently, a new approach to electronic structure calculations of strongly correlated materials
involving $d-$ or $f-$ orbitals, has been developed. This approach,
dubbed ``LDA+DMFT'', combines the dynamical
mean-field theory (DMFT) \cite{DMFT}
of correlated electron models with DFT-LDA calculations\cite{LDA+DMFT}.
It is also a Green's function technique, but -- unlike GWA -- it
does not treat
the Coulomb interaction from first principles. Instead,
an effective Hamiltonian involving Hubbard-like
interaction {\it parameters} in the restricted subset of correlated orbitals is used as a starting point.
It is thus necessary to introduce a ``double-counting''
correction term. The strength of DMFT however is that the
onsite electronic interactions are
treated to all orders, by using a mapping onto a self-consistent quantum impurity problem.
DMFT has led to remarkable advances on electronic structure calculations
of materials in which the Mott phenomenon
or the formation of local moments play a key role.
This is the case, e.g. for the satellite structure in Ni, which has recently
been shown to be correctly described by LDA+DMFT \cite{lichtenstein_ni}.

The aim of this letter is to take a new step towards a first-principles electronic structure
calculation method for strongly correlated materials. We propose a scheme in which
the GW treatment of the screened Coulomb interaction and exchange self-energy is combined
with a DMFT calculation for the
onsite components of these two quantities,
in a self-consistent manner \cite{kotliar_lecture}. The frequency-dependence of the
onsite effective interaction
(or polarization) actually requires an extended DMFT scheme (E-DMFT), as introduced in earlier work
in a model context for both the charge and spin channels
\cite{si_smith_edmft_1996,kajueter_1995,kajueter_phd,
sengupta_georges_metallic_sg}.
This combined GW~$+$~(E)DMFT scheme
does not make use of Hubbard-like interaction parameters and bypasses the need for a
double-counting correction when implemented in a self-consistent dynamical manner.
In fact, using LDA is in principle no longer necessary within
such a self-consistent implementation.
In this work however, we implement a simplified version of this scheme on the example of
ferromagnetic nickel, which serves as a test for the feasibility of realistic calculations
using this approach.

We consider the Hamiltonian for electrons in a solid
interacting via the Coulomb potential
$V(\vr-\vr')=e^2/|\vr-\vr'|$.
The general strategy of our approach is to construct a functional
of the
one-electron Green's function
$G(\vr,\vr';\tau-\tau') \equiv - \bra T_\tau\,\psi(\vr,\tau)\psi^\dagger(\vr',\tau')\ket$
and the screened Coulomb interaction $W$ \cite{almbladh_functionals,
chitra_bk}.
Here, $T_\tau$ denotes the time ordering operator in imaginary
time and $\psi(\vr,\tau)$ [$\psi^\dagger(\vr,\tau)$] the
annihilation [creation] operator of an electron at point $\vr$ at
time $\tau$.
The screened Coulomb interaction is
defined using the (connected) density-density
response function:
$\chi(\vr,\vr';\tau-\tau')\equiv \bra
T_\tau[\hro(\vr,\tau)-n(\vr)][\hro(\vr',\tau')-n(\vr')]\ket$
\begin{eqnarray}\label{eq:def_W}
W(\vr,\vr';\iomn) &=& V (\vr-\vr') -
\\
\nonumber
&-& \int d\vr_1 d\vr_2 V(\vr-\vr_1)\chi(\vr_1,\vr_2;\iomn)V(\vr_2-\vr')
\end{eqnarray}
Following \cite{almbladh_functionals} and \cite{chitra_bk}
we introduce the
free-energy functional
(which generalizes the Luttinger-Ward
construction)
\begin{eqnarray}\label{eq:def_gamma}
\nonumber
\Gamma[G,W]\,\,=\,\,\T\ln G -\T [(\GH^{-1}-G^{-1})G]\\
-\frac{1}{2}\T\ln W
+ \frac{1}{2}\T [(V^{-1}-W^{-1})W] + \Psi [G,W]
\end{eqnarray}
In this expression
$\GH^{-1}=\iomn+\mu+\nabla^2/2-v_c-v_H$ is the bare Green's function of
the solid including the Hartree potential
$ v_H(\vr)\equiv\int d\vr' V(\vr-\vr') n(\vr')$.
$\Psi$ is the contribution to the functional
due to electronic correlations beyond Hartree. It corresponds to the sum of skeleton
diagrams which are irreducible with respect to both the one-electron propagator and the
interaction.

A formal construction of this functional
can be given (following \cite{chitra_bk})
by making a Hubbard-Stratonovich transformation, using
auxiliary bosonic fields $\phi(\vr,\tau)$ conjugate to the density fluctuations
$[\psi^\dagger(\vr,\tau)\psi(\vr,\tau)-n(\vr)]$. The effective interaction precisely
corresponds to the boson correlator:
$W(\vr,\vr',\tau-\tau')=\bra T_\tau\phi(\vr,\tau)\phi(\vr',\tau')\ket$.
The functional $\Gamma$ is then constructed by a Legendre transformation with respect to
both $G$ and $W$. A formal expression of the correlation functional $\Psi[G,W]$
(generalizing the Luttinger-Ward $\Phi[G]$) can be given, using an integration over
a coupling constant parameter $\alpha$ between the bosonic and fermionic variables:
$\Psi[G,W] = i\,\int_0^1 d\a \int d\vr\,d\tau
\bra \phi(\vr,\tau)[\psi^\dagger(\vr,\tau)\psi(\vr,\tau)-n(\vr)]\ket$.
The GW approximation retains only the first order contribution to this functional
in the $\a$-expansion, corresponding to the exchange diagram
$\Psi_{GW} = - \frac{1}{2}\mbox{Tr} GWG$
\cite{almbladh_functionals}.

The equilibrium state of the system corresponds to a stationary point of the
functional $\Gamma$, which leads to the identification of the
exchange and correlation self-energy $\Sigma^{xc}$ and of the polarization
operator $P$:
\begin{eqnarray}\label{eq:def_sigma_P}\nonumber
\frac{\d\Gamma}{\d G}=0 \Rightarrow G^{-1}=\GH^{-1}-\Sigma^{xc}\,\,\,,\,\,\,
\Sigma^{xc} = \frac{\d\Psi}{\d G}\\
\frac{\d\Gamma}{\d W}=0 \Rightarrow W^{-1}=V^{-1}-P\,\,\,,\,\,\,
P = -2 \frac{\d\Psi}{\d W}
\end{eqnarray}
In the (self-consistent) GW approximation: $\Sigma^{xc}_{GW}=-G~W$ and
$P_{GW}= G~G$ (the signs result from the use of the Matsubara formalism).

In order to proceed further, we need to specify a basis set.
One-particle quantities like $G$ or $\Sigma$ are represented as
$G(\vr,\vr',\iomn)=
\sum_{LL'\vR\vR'} \phi_L^\vR(\vr)
G_{LL'}^{\vR\vR'}(\iomn) \phi_{L'}^{\vR'}(\vr')^* =
\sum_{LL'\vk} \phi_L^\vk(\vr)
G_{LL'}(\vk,\iomn) \phi_{L'}^\vk(\vr')^*$
where $\phi$ are localized basis functions (e.g. LMTO's) \cite{lmto},
centered at an atomic
position R (and for simplicity assumed to be orthogonal).
Two-particle
quantities such as P or W are represented as
 $W(\vr,\vr',i\nu_n)=\sum_{\a\b\vR\vR'} B_\a^\vR(\vr)
 W_{\a\b}^{\vR\vR'}(i\nu_n) B_\b^{\vR'}(\vr')^*$.
Here B's are linear combinations of $\phi \phi$ and form an orthonormal set \cite{ferdi_gw_review}.
Note that the set $\phi \phi$ is in general overcomplete so that the number of
B's is smaller or equal to the number of $\phi \phi$. Matrix elements in
products of LMTOs are then given by
\begin{equation}\label{trafo}
W_{L_1L_2L_3L_4}^{\vR\vR'}\equiv
\langle\phi_{L_1}^{\vR}\phi_{L_2}^{\vR}|W|\phi_{L_3}^{\vR'}\phi_{L_4}^{\vR'}\rangle =
\sum_{\a\b}O_{L_1L_2}^\a W_{\a\b}^{\vR\vR'} O_{L_3L_4}^{\b *}
\end{equation}
with the overlap
matrix $O_{L_1L_2}^\a\equiv\langle\phi_{L_1}\phi_{L_2}|B^\a\rangle$.
We note that in general we cannot obtain $W_{\alpha\beta}$
from $W_{L_1L_2L_3L_4}$, while the converse is true.

The functionals $\Gamma[G,W]$ and $\Psi[G,W]$ can thus be viewed as functionals
of the matrix elements
$G_{L_1L_2}^{\vR\vR'}(\iomn)$ and $W_{\a\b}^{\vR\vR'}(\iomn)$.
The main idea behind the present work is that the dependence of the
$\Psi$-functional upon the
{\it off-site} components ($\vR\neq\vR'$) of
$G^{\vR\vR'}$ and $W^{\vR\vR'}$ can be treated within the GW approximation,
while the dependence on the
{\it onsite} components
($\vR=\vR'$) requires a more accurate treatment.
For strongly correlated systems, the
{\it onsite} effective interaction will enter the
strong-coupling regime in which
an RPA treatment is insufficient. We thus approximate the functional
$\Psi$ as:
\begin{equation}\label{Psi}
\Psi\,=\,\Psi_{GW}^{\rm{non-loc}}[\GRRp,\WRRp]+\Psi_{\imp}[\GRR,\WRR]
\end{equation}
In this expression, the first term corresponds to the GW-functional (written in the specified
basis set) and restricted to
off-site components of $G$ and $W$ (i.e associated with
distinct spheres $\vR\neq\vR'$), namely:
\begin{eqnarray}\nonumber
\Psi_{GW}^{\rm{non-loc}}&=
-\frac{1}{2}\int d\tau \sum_{L_1\cdots L_2'}
\sum_{\vR\neq\vR'}
\nonumber
\\
& \GRRp_{L_1L'_1}(\tau)\WRRp_{L_1L_2L'_1L'_2}(\tau)G^{\vR'\vR}_{L_2'L_2}(-\tau)
\end{eqnarray}
with $\WRRp_{L_1L_2L'_1L'_2}$ given by (\ref{trafo}).
Let us note that $\Psi_{GW}^{\rm{non-loc}}$ can also be written as the difference between the
complete GW-functional, and the contributions from the
onsite components:
$\Psi_{GW}^{\rm{non-loc}}=\Psi_{GW}-\Psi_{GW}^{\rm{loc}}[\GRR,\WRR]$.
All the dependence on these onsite components is gathered into $\Psi_{imp}$.
Following (extended) DMFT, this onsite part of the functional is generated
\footnote{$\Psi_{imp}$ is related to the functional $\Gamma_{imp}[G_{imp},W_{imp}]$ associated
with this local problem by an expression identical to (\ref{eq:def_gamma}), with
${\cal G}$ replacing $\GH$ and ${\cal U}$ replacing $V$.}
from a local {\it quantum impurity problem} (defined on a single atomic site),
with effective action:
\begin{eqnarray}\label{eq:action}\label{Simp}
&S&=\int d\tau 
d\tau' \left[ -
\sum 
c^{\dagger}_{L}(\tau) {\cal G}^{-1}_{LL'}(\tau-\tau')c_{L'}(\tau')
\right.
\\ \nonumber
&+&\frac{1}{2}
\left.
\sum 
:c^\dagger_{L_1}(\tau)c_{L_2}(\tau):\cU_{L_1L_2L_3L_4}(\tau-\tau')
:c^\dagger_{L_3}(\tau')c_{L_4}(\tau'): \right]
\hskip-1cm
\end{eqnarray}
where the sums run over all orbital indices $L$.
In this expression, $c_L^+$ is a creation operator associated with orbital $L$
on a given sphere, and the double dots denote normal ordering (taking care of
Hartree terms).
This can be viewed as a representability assumption, namely that
the local components of $G$ and $W$ can be obtained from (\ref{Simp})
with suitably chosen values of the auxiliary (Weiss) functions
$\cG$ and $\cU$. This is formally analogous to the Kohn-Sham representation of the
local density in a solid. This construction defines the
(frequency-dependent) Hubbard interactions $\cU_{L_1L_2L_3L_4}(\omega)$,
for a specific material, in a unique manner (for a given basis set).
Note that $\cU_{L_1L_2L_3L_4}$ must correspond
to an interaction matrix $\cU_{\a\b}$ in the two-particle basis $B^\a$
via a transformation identical to (\ref{trafo}).
Taking derivatives of (\ref{Psi}) as in (\ref{eq:def_sigma_P}) it is seen
that the complete self-energy and polarization operators read:
\begin{eqnarray}\label{Sig_a}
\Sigma^{xc}(\vk,\iomn)_{LL'} &=& \Sigma_{GW}^{xc}(\vk,\iomn)_{LL'}
\\
&-& \sum_\vk \Sigma_{GW}^{xc}(\vk,\iomn)_{LL'}
+ [\Sigma^{xc}_{imp}(\iomn)]_{LL'}
\nonumber
\\
\label{P_a}
P(\vq,\inun)_{\a\b} &=& P^{GW}(\vq,\inun)_{\a\b}
\\
&-& \sum_\vq P^{GW}(\vq,\inun)_{\a\b}+ P^{imp}(\inun)_{\a\b}
\nonumber
\end{eqnarray}
The meaning of (\ref{Sig_a}) is transparent:
the off-site part of the self-energy is taken from
the GW approximation, whereas the onsite part is calculated
to all orders
from the dynamical impurity model.
This treatment thus goes beyond usual E-DMFT, where
the lattice self-energy and polarization are just taken
to be their impurity counterparts.
The second term in
(\ref{Sig_a}) substracts the onsite component of the GW
self-energy thus avoiding double counting.
As explained below, at self-consistency this term can be rewritten as:
\begin{equation}\label{Sig_correction}
\sum_\vk \Sigma_{GW}^{xc}(\tau)_{LL'}= - \sum_{L_1L_1'}
W^{imp}_{LL_1L'L'_1}(\tau) G_{L'_1L_1}(\tau)
\end{equation}
(where, again, $W^{imp}_{LL_1L'L'_1}$ is related to $W^{imp}_{\a\b}$
by an equation of the type (\ref{trafo})) so that
it precisely substracts the contribution of the GW diagram to
the impurity self-energy. Similar considerations apply
to the polarization operator.

We now outline the iterative loop which determines $\cG$ and $\cU$ self-consistently
(and, eventually, the full self-energy and polarization operator):
\begin{itemize}
\item The impurity problem (\ref{Simp}) is solved, for a given choice of $\cG_{LL'}$ and
$\cU_{\a\b}$: the ``impurity'' Green's function
$G_{imp}^{LL'}\equiv - \langle T_\tau c_L(\tau)c^+_{L'}(\tau')\rangle_S$
is calculated, together with the impurity self-energy
$\Sigma^{xc}_{imp}\equiv\delta\Psi_{imp}/\delta G_{imp}=\cG^{-1}-G_{imp}^{-1}$.
The two-particle correlation function
$\chi_{L_1L_2L_3L_4}=\langle :c^\dagger_{L_1}(\tau)c_{L_2}(\tau):
:c^\dagger_{L_3}(\tau')c_{L_4}(\tau'):\rangle_S$ must also be evaluated.
\item
The impurity effective interaction is constructed as follows:
\begin{equation}\label{eff_inter}
W_{imp}^{\a\b} = \cU_{\a\b} -
\sum_{L_1\cdots L_4}\sum_{\gamma\delta} \cU_{\a\gamma}
O^{\gamma}_{L_1L_2}
\chi_{L_1L_2L_3L_4} [O^{\delta}_{L_3L_4}]^* \cU_{\delta\b}
\end{equation}
Here all quantities are evaluated at the same frequency%
\footnote{Note that $\chi_{L_1...L_4}$ does {\bf not} denote the matrix
element $<L_1L_2|\chi|L_3L_4>$, but is rather defined by
$\chi(r,r') = \sum_{L_1..L_4}
\phi^{\ast}_{L_1}(r)  \phi^{\ast}_{L_2}(r)  \chi_{L_1...L_4} \phi_{L_3}(r')  \phi_{L_4}(r') $.}.
The polarization operator of the impurity problem is then obtained as:
$P_{imp}\equiv -2\delta\Psi_{imp}/\delta W_{imp} = \cU^{-1}-W_{imp}^{-1}$,
where the matrix inversions are performed in the two-particle basis $B^\a$.
\item
From Eqs.~(\ref{Sig_a}) and (\ref{P_a})
the full $\vk$-dependent Green's function $G(\vk,\iomn)$ and
effective interaction $W(\vq,\inun)$ can be constructed. The self-consistency condition
is obtained, as in the usual DMFT context, by requiring that the onsite
components of these quantities coincide with $G_{imp}$ and $W_{imp}$. In practice, this
is done by computing the onsite quantities
\begin{eqnarray}\label{Glocal}
G_{loc}(\iomn) &=& \sum_\vk [\GH^{-1}(\vk,\iomn) - \Sigma^{xc}(\vk,\iomn) ]^{-1}
\\
\label{Wlocal}
W_{loc}(\inun) &=& \sum_\vq [V_{\vq}^{-1} - P(\vq,\inun)]^{-1}
\end{eqnarray}
and using them to update the
Weiss dynamical mean field ${\cal G}$
and the impurity model interaction ${\cal U}$ according to:
\begin{eqnarray}\label{update}
\cG^{-1} = G_{loc}^{-1} + \Sigma_{imp}
\\
\cU^{-1} = W_{loc}^{-1} + P_{imp}
\end{eqnarray}
\end{itemize}
This cycle is iterated until self-consistency for
$\cG$ and $\cU$ is obtained (as well as on $G$, $W$, $\Sigma^{xc}$ and $P$).
Eventually, self-consistency over the local electronic density can also
be implemented,
(in a similar way as in LDA+DMFT \cite{savrasov_pu,savrasov_functional})
by recalculating $\rho(\vec{r})$ from the Green's function at
the end of the convergence cycle above, and constructing an updated Hartree potential.
This new density is used as an input of a new GW calculation,
and convergence over this external loop must be reached.
While implementing self-consistency within the GWA is known to yield
unsatisfactory spectra \cite{holm}, we expect a more favorable
situation in the proposed GW$+$DMFT scheme since
part of the interaction effects
are treated to all orders.

\begin{figure}
\centerline{\includegraphics[width=8cm]{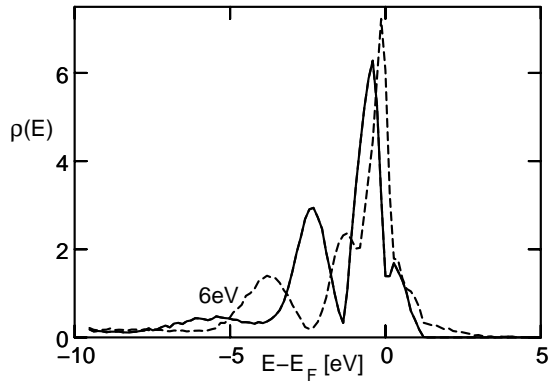}}
\caption{\label{dos}
Partial density of states of d-orbitals of
nickel (solid [dashed] lines give the majority [minority]
spin contribution) as obtained from the
combination of GW and DMFT (see text).
For comparison with LDA and LDA+DMFT results see
\cite{lichtenstein_ni}, 
for experimental spectra see \cite{maartenson}.
}
\end{figure}

The practical implementation of the proposed approach in a fully dynamical
and self-consistent manner is an ambitious task, which we regard as a
major challenge for future research.
Here, we only demonstrate the feasibility and potential of the approach
within a simplified implementation, which we apply to the electronic
structure of Nickel.
The main simplifications made are:
(i) The DMFT local treatment is applied only to the
$d$-orbitals,
(ii) the GW calculation is done only once, in the form \cite{ferdi_gw_review}:
$\Sigma^{xc}_{GW} = G_{LDA}\cdot W[G_{LDA}]$, from which the
non-local part of the self-energy is obtained,
(iii) we replace the dynamical impurity problem
by its static limit, solving the impurity model (\ref{Simp})
for a frequency-independent $\cU=\cU(\omega=0)$.
Instead of the Hartree Hamiltonian 
we start from
a one-electron Hamiltonian in the form:
$H_{LDA} - V_{xc,\sigma}^{nonlocal} - \frac{1}{2} \mbox{\bf Tr}
\Sigma_{\sigma}^{imp}(0)$.
The non-local part of this Hamiltonian coincides with that of 
the Hartree Hamiltonian
while its local part is derived from LDA, with a double-counting
correction of the form proposed in
\cite{lichtenstein_ni} in the DMFT context.
With this choice the self-consistency condition
(\ref{Glocal}) reads:
\begin{eqnarray}\label{Glocal2}
G_{loc}^{\sigma}(\iomn) &=& \sum_\vk [
\GH^{-1}(\vk,\iomn)
- (\Sigma^{xc}_{GW})_{non-loc}
\\ \nonumber
&-&(\Sigma_{imp,\sigma} - \frac{1}{2} \mbox{\bf Tr}_{\sigma}
\Sigma_{imp,\sigma}(0) + V_{xc}^{loc})\,]^{-1}
\end{eqnarray}
We have performed finite temperature GW and LDA+DMFT
calculations (within the LMTO-ASA\cite{lmto} with
29 irreducible $\vk$-points) for ferromagnetic nickel
(lattice constant 6.654 a.u.), using 4s4p3d4f
states, at the Matsubara frequencies
$\iomn$ corresponding to $T=630 K$, just
below the Curie temperature.
The resulting self-energies are inserted into
Eq.~(\ref{Glocal2}), which is then used to calculate
a new Weiss field according to
(\ref{update}).
The Green's function $G^{\sigma}_{loc}(\tau)$ is recalculated
from the impurity effective action by 
QMC and analytically continued using the
Maximum Entropy algorithm.
The resulting spectral function is plotted in Fig.(\ref{dos}).
Comparison with the LDA+DMFT results in \cite{lichtenstein_ni}
shows that the good description of the satellite structure,
exchange splitting and band narrowing is indeed retained
within the (simplified) GW+DMFT scheme.

\begin{figure}
\centerline{\includegraphics[width=8cm]{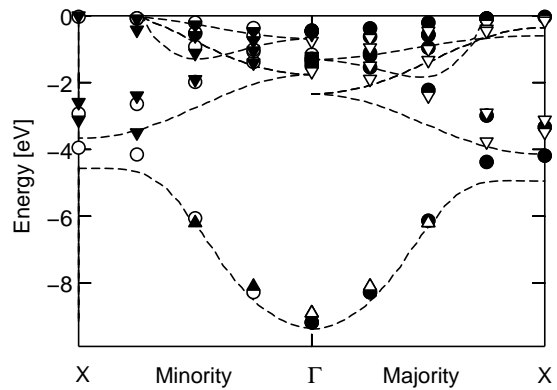}}
\caption{\label{bands}
Band structure of Ni (minority and majority spin) as obtained from
the linearization procedure of the GW+DMFT self-energy
described in the text (dots) in comparison to the
LDA band structure (dashed lines) and the experimental
data of \cite{buenemann} (triangles down) and \cite{maartenson}
(triangles up).}
\end{figure}

We have also calculated the quasiparticle band structure,
from the 
poles of (\ref{Glocal2}), after linearization of $\Sigma(\vk,\iomn)$
around the Fermi level
\footnote{Note however that this linearization is no
longer meaningful at energies far away from the Fermi
level. We therefore use the unrenormalized value for
the quasi-particle residue for the s-band ($Z_{s}=1$).}.
Fig.~(\ref{bands}) shows a comparison of GW+DMFT with the
LDA and experimental band structure. It is seen that GW+DMFT correctly
yields the bandwidth reduction compared to the (too large)
LDA value and renormalizes the bands in a
($\vk$-dependent) manner.

We now discuss further the simplifications made in our implementation.
Because of the static approximation (iii), we could not implement self-consistency
on $W_{loc}$ (Eq. (\ref{Wlocal})).
We chose the value of $\cU (\omega=0)$ ($\simeq 3.2 eV$) by calculating
the correlation function $\chi$ and ensuring that
Eq. (\ref{eff_inter}) is fulfilled at $\omega=0$, given the GW value for
$W_{loc}(\omega=0)$ ($\simeq 2.2 eV$ for Nickel \cite{ferdi_W}).
This procedure emphasizes the low-frequency, screened value, of the effective
interaction. Obviously, the resulting impurity self-energy
$\Sigma_{imp}$ is then much smaller than the local component of the
GW self-energy (or than $V_{xc}^{loc}$), especially at high frequencies.
It is thus essential to choose the second term in (\ref{Sig_a}) to
be the onsite component of the GW self-energy rather than the
r.h.s of Eq.~(\ref{Sig_correction}). For the same reason, we
included $V_{xc}^{loc}$ in Eq.(\ref{Glocal2}) (or, said differently, we
implemented a mixed scheme which starts from the LDA Hamiltonian for the
local part, and thus still involves a double-counting correction).
We expect that these limitations can be overcome in a self-consistent
implementation with a frequency-dependent $\cU(\omega)$
(hence fulfiling Eq.~(\ref{Sig_correction})). In practice, it might be
sufficient to replace the local part of the GW self-energy
by $\Sigma_{imp}$ for correlated orbitals only. Alternatively, a
downfolding procedure could be used.

In conclusion, we have proposed an {\it ab initio}
dynamical mean field approach for calculating the
electronic structure of strongly correlated materials,
which combines GW and DMFT.
The scheme aims at avoiding
the conceptual problems inherent to ``LDA+DMFT'' methods, such as
double counting corrections and
the use of Hubbard parameters assigned to correlated orbitals.
A full practical implementation of the GW$+$DMFT scheme is a major goal
for future research, which requires further work on impurity models
with frequency-dependent
interaction parameters \cite{motome_dynamical_QMC,jarrell_holstein,sun}
as well as studies of various possible self-consistency schemes.

\begin{acknowledgments}
During completion of this work, we learnt about Ref.~\cite{sun}
in which a GW correction to the E-DMFT scheme has been successfully
implemented, in a dynamical manner, for a one-dimensional
extended Hubbard model. We thank G.~Kotliar for
providing a copy of this work prior to publication.
We are grateful, for comments and helpful discussions, to:
S. Florens, G. Kotliar, P. Sun and to A. Lichtenstein
(who also shared with us his QMC code).
This work has benefitted from the hospitality of the
MPI-FKF Stuttgart (for which we thank O. K. Andersen) and of
the KITP-UCSB (under NSF grant PHY99-07949).
It has been supported by a Marie Curie Fellowship of the
EC Programme ``Improving Human Potential'' under
contract number HPMF CT 2000-00658
and by a grant of supercomputing time at IDRIS (CNRS, Orsay).

We dedicate this paper to the memory of Lars Hedin.
We were very fortunate to receive his encouragements
to pursue the present work.

\end{acknowledgments}


\begin{thebibliography}{21}
\expandafter\ifx\csname natexlab\endcsname\relax\def\natexlab#1{#1}\fi
\expandafter\ifx\csname bibnamefont\endcsname\relax
  \def\bibnamefont#1{#1}\fi
\expandafter\ifx\csname bibfnamefont\endcsname\relax
  \def\bibfnamefont#1{#1}\fi
\expandafter\ifx\csname citenamefont\endcsname\relax
  \def\citenamefont#1{#1}\fi
\expandafter\ifx\csname url\endcsname\relax
  \def\url#1{\texttt{#1}}\fi
\expandafter\ifx\csname urlprefix\endcsname\relax\def\urlprefix{URL }\fi
\providecommand{\bibinfo}[2]{#2}
\providecommand{\eprint}[2][]{\url{#2}}

\bibitem[{\citenamefont{Hedin}(1965)}]{hedin_gw}
\bibinfo{author}{\bibfnamefont{L.}~\bibnamefont{Hedin}},
  \bibinfo{journal}{Phys. Rev.} \textbf{\bibinfo{volume}{139}}
  (\bibinfo{year}{1965}).

\bibitem[{\citenamefont{Aryasetiawan and Gunnarsson}(1998)}]{ferdi_gw_review}
\bibinfo{author}{\bibfnamefont{F.}~\bibnamefont{Aryasetiawan}}
  \bibnamefont{and}
  \bibinfo{author}{\bibfnamefont{O.}~\bibnamefont{Gunnarsson}},
  \bibinfo{journal}{Rep. Prog. Phys.} \textbf{\bibinfo{volume}{61}},
  \bibinfo{pages}{237} (\bibinfo{year}{1998}).

\bibitem[{\citenamefont{Onida et~al.}(2002)\citenamefont{Onida, Reining, and
  Rubio}}]{reining_gw_rmp}
\bibinfo{author}{\bibfnamefont{G.}~\bibnamefont{Onida}},
  \bibinfo{author}{\bibfnamefont{L.}~\bibnamefont{Reining}}, \bibnamefont{and}
  \bibinfo{author}{\bibfnamefont{A.}~\bibnamefont{Rubio}},
  \bibinfo{journal}{Rev. Mod. Phys.} \textbf{\bibinfo{volume}{74}},
  \bibinfo{pages}{601} (\bibinfo{year}{2002}).

\bibitem[{\citenamefont{Aryasetiawan}(1992)}]{ferdi_gw_nickel}
\bibinfo{author}{\bibfnamefont{F.}~\bibnamefont{Aryasetiawan}},
  \bibinfo{journal}{Phys. Rev. B} \textbf{\bibinfo{volume}{46}},
  \bibinfo{pages}{13051} (\bibinfo{year}{1992}).

\bibitem[{DMF()}]{DMFT}
  \bibinfo{note}{For reviews see: A. Georges {\it et al.}, Rev. Mod.
  Phys. {\bf 68}, 13 (1996); T. Pruschke {\it et al.}, Adv. Phys. {\bf 42}, 187
  (1995)}.

\bibitem[{LDA()}]{LDA+DMFT}
  \bibinfo{note}{For recent reviews see: {\it "Strong Coulomb
  Correlations in Electronic Structure Calculations"} (Advances in Condensed
  Matter Science), Edited by V.~Anisimov. Gordon and Breach (2001);
  }.

\bibitem[{\citenamefont{Lichtenstein et~al.}(2001)\citenamefont{Lichtenstein,
  Katsnelson, and Kotliar}}]{lichtenstein_ni}
\bibinfo{author}{\bibfnamefont{A.~I.} \bibnamefont{Lichtenstein}},
  \bibinfo{author}{\bibfnamefont{M.~I.} \bibnamefont{Katsnelson}},
  \bibnamefont{and} \bibinfo{author}{\bibfnamefont{G.}~\bibnamefont{Kotliar}},
  \bibinfo{journal}{Phys. Rev. Lett.} \textbf{\bibinfo{volume}{87}},
  \bibinfo{pages}{067205} (\bibinfo{year}{2001}).

\bibitem[{\citenamefont{Savrasov and Kotliar}(2001)}]{kotliar_lecture}
  \bibinfo{note}{For related ideas, see: G. Kotliar and S. Savrasov in
   {\it New Theoretical Approaches to Strongly Correlated Systems},
   Ed. by A. M. Tsvelik (2001) Kluwer Acad. Publ.
   (and the updated version: cond-mat/0208241)
   }.

\bibitem[{\citenamefont{Si and Smith}(1996)}]{si_smith_edmft_1996}
\bibinfo{author}{\bibfnamefont{Q.}~\bibnamefont{Si}} \bibnamefont{and}
  \bibinfo{author}{\bibfnamefont{J.~L.} \bibnamefont{Smith}},
  \bibinfo{journal}{Phys. Rev. Lett.} \textbf{\bibinfo{volume}{77}},
  \bibinfo{pages}{3391} (\bibinfo{year}{1996}).

\bibitem[{\citenamefont{Kotliar and Kajueter}(1995)}]{kajueter_1995}
\bibinfo{author}{\bibfnamefont{G.}~\bibnamefont{Kotliar}} \bibnamefont{and}
  \bibinfo{author}{\bibfnamefont{H.}~\bibnamefont{Kajueter}},
  \bibinfo{journal}{unpublished}  (\bibinfo{year}{1995}).

\bibitem[{\citenamefont{Kajueter}(1996)}]{kajueter_phd}
\bibinfo{author}{\bibfnamefont{H.}~\bibnamefont{Kajueter}},
  \bibinfo{journal}{PhD thesis, Rutgers University}  (\bibinfo{year}{1996}).


\bibitem[{\citenamefont{Sengupta and
  Georges}(1995)}]{sengupta_georges_metallic_sg}
\bibinfo{author}{\bibfnamefont{A.~M.}~\bibnamefont{Sengupta}} \bibnamefont{and}
  \bibinfo{author}{\bibfnamefont{A.}~\bibnamefont{Georges}},
  \bibinfo{journal}{Phys. Rev. B} \textbf{\bibinfo{volume}{52}},
  \bibinfo{pages}{10295} (\bibinfo{year}{1995}).

\bibitem[{\citenamefont{Almbladh et~al.}(1999)\citenamefont{Almbladh, von
  Barth, and van Leeuwen}}]{almbladh_functionals}
\bibinfo{author}{\bibfnamefont{C.~O.} \bibnamefont{Almbladh}},
  \bibinfo{author}{\bibfnamefont{U.}~\bibnamefont{von Barth}},
  \bibnamefont{and} \bibinfo{author}{\bibfnamefont{R.}~\bibnamefont{van
  Leeuwen}}, \bibinfo{journal}{Int. J. Mod. Phys. B}
  \textbf{\bibinfo{volume}{13}}, \bibinfo{pages}{535} (\bibinfo{year}{1999}).

\bibitem[{\citenamefont{Chitra and Kotliar}(2001)}]{chitra_bk}
\bibinfo{author}{\bibfnamefont{R.}~\bibnamefont{Chitra}} \bibnamefont{and}
  \bibinfo{author}{\bibfnamefont{G.}~\bibnamefont{Kotliar}},
  \bibinfo{journal}{Phys. Rev. B} \textbf{\bibinfo{volume}{63}},
  \bibinfo{pages}{115110} (\bibinfo{year}{2001}).


\bibitem[{\citenamefont{Andersen}(1975)}]{lmto}
\bibinfo{author}{\bibfnamefont{O.~K.} \bibnamefont{Andersen}},
  \bibinfo{journal}{Phys.~Rev.~B} \textbf{\bibinfo{volume}{12}},
  \bibinfo{pages}{3060} (\bibinfo{year}{1975}).

\bibitem[{\citenamefont{Savrasov and Kotliar}(2001)}]{savrasov_pu}
\bibinfo{author}{\bibfnamefont{S.}~\bibnamefont{Savrasov}} \bibnamefont{and}
  \bibinfo{author}{\bibfnamefont{G.}~\bibnamefont{Kotliar}},
  \bibinfo{journal}{cond-mat/0106308}  (\bibinfo{year}{2001}).

\bibitem[{\citenamefont{Savrasov et~al.}(2000)\citenamefont{Savrasov, Kotliar,
  and Abrahams}}]{savrasov_functional}
\bibinfo{author}{\bibfnamefont{S.}~\bibnamefont{Savrasov}},
  \bibinfo{author}{\bibfnamefont{G.}~\bibnamefont{Kotliar}}, \bibnamefont{and}
  \bibinfo{author}{\bibfnamefont{E.}~\bibnamefont{Abrahams}},
  \bibinfo{journal}{Nature} \textbf{\bibinfo{volume}{410}},
  \bibinfo{pages}{793} (\bibinfo{year}{2000}).

\bibitem[{\citenamefont{Holm and von Barth}(1998)}]{holm}
\bibinfo{author}{\bibfnamefont{B.}~\bibnamefont{Holm}}
  \bibnamefont{and}
  \bibinfo{author}{\bibfnamefont{U.}~\bibnamefont{von Barth}},
  \bibinfo{journal}{Phys. Rev. B} \textbf{\bibinfo{volume}{57}},
  \bibinfo{pages}{2108} (\bibinfo{year}{1998}).

\bibitem[{\citenamefont{Springer and Aryasetiawan}(1998)}]{ferdi_W}
\bibinfo{author}{\bibfnamefont{M.}~\bibnamefont{Springer}} \bibnamefont{and}
  \bibinfo{author}{\bibfnamefont{F.}~\bibnamefont{Aryasetiawan}},
  \bibinfo{journal}{Phys. Rev. B} \textbf{\bibinfo{volume}{57}},
  \bibinfo{pages}{4364} (\bibinfo{year}{1998}).

\bibitem[{\citenamefont{B{\"u}nemann et~al.}(2002)}]{buenemann}
\bibinfo{author}{\bibfnamefont{J.}~\bibnamefont{B{\"u}nemann et~al.}}
  (\bibinfo{year}{2002}), \bibinfo{note}{cond-mat/0204142}.

\bibitem[{\citenamefont{M{\aa}rtensson and Nilsson}(1984)}]{maartenson}
\bibinfo{author}{\bibfnamefont{H.}~\bibnamefont{M{\aa}rtensson}}
  \bibnamefont{and}
  \bibinfo{author}{\bibfnamefont{P.~O.}~\bibnamefont{Nilsson}},
  \bibinfo{journal}{Phys. Rev. B} \textbf{\bibinfo{volume}{30}},
  \bibinfo{pages}{3047} (\bibinfo{year}{1984}).

\bibitem[{\citenamefont{Motome and Kotliar}(2000)}]{motome_dynamical_QMC}
\bibinfo{author}{\bibfnamefont{Y.}~\bibnamefont{Motome}} \bibnamefont{and}
  \bibinfo{author}{\bibfnamefont{G.}~\bibnamefont{Kotliar}},
  \bibinfo{journal}{Phys. Rev. B} \textbf{\bibinfo{volume}{62}},
  \bibinfo{pages}{12800} (\bibinfo{year}{2000}).

\bibitem[{\citenamefont{Freericks et~al.}(1993)\citenamefont{Freericks,
  Jarrell, and Scalapino}}]{jarrell_holstein}
\bibinfo{author}{\bibfnamefont{J.~K.} \bibnamefont{Freericks}},
  \bibinfo{author}{\bibfnamefont{M.}~\bibnamefont{Jarrell}}, \bibnamefont{and}
  \bibinfo{author}{\bibfnamefont{D.~J.} \bibnamefont{Scalapino}},
  \bibinfo{journal}{Phys. Rev. B} \textbf{\bibinfo{volume}{48}},
  \bibinfo{pages}{6302} (\bibinfo{year}{1993}).

\bibitem[{\citenamefont{Sun and Kotliar}(2002)}]{sun}
\bibinfo{author}{\bibfnamefont{P.}~\bibnamefont{Sun}} \bibnamefont{and}
  \bibinfo{author}{\bibfnamefont{G.}~\bibnamefont{Kotliar}}
  (\bibinfo{year}{2002}), \bibinfo{note}{cond-mat/0205522}.

\end{thebibliography}
\end{document}